\newcommand{\be}{\begin{equation}}
\newcommand{\ee}{\end{equation}}
\newcommand{\ba}{\begin{array}}
\newcommand{\ea}{\end{array}}
\newcommand{\beqn}{\begin{eqnarray}}
\newcommand{\eeqn}{\end{eqnarray}}

\newcommand{\zero}{\setcounter{equation}{0} \hfill }

\documentclass[12pt]{article}
\usepackage[dvips]{graphicx}

\begin{document}
\title{Banking retail consumer finance data generator - credit scoring data repository}
\author{Karol Przanowski\\
email: \texttt{kprzan@interia.pl} \\
url: \texttt{http://kprzan.w.interia.pl}
}
\date{}
\maketitle
\begin{abstract}
This paper presents two cases of random banking data generators based on migration matrices and scoring rules. The banking data generator is a new hope in researches of finding the proving method of comparisons of various credit scoring techniques. There is analyzed the influence of one cyclic macro--economic variable on stability in the time account and client characteristics. Data are very useful for various analyses to understand in the better way the complexity of the banking processes and also for students and their researches. There are presented very interesting conclusions for crisis behavior, namely that if a crisis is impacted by many factors, both customer characteristics: application and behavioral; then there is very difficult to indicate these factors in the typical scoring analysis and the crisis is everywhere, in every kind of risk reports.    
\end{abstract}
\setcounter{section}{0}
\setcounter{page}{1}
\def\gw{\vskip0.5cm
\centerline{***}
\vskip0.5cm}

\centerline{{\bf Key words:} credit scoring, crisis, banking data generator, retail portfolio.}

\newpage

\tableofcontents

\newpage
\section{Introduction \label{intro}}
\zero
\par

Currently predictive models and especially credit scoring models are very popular in management of banking processes~\cite{huangimportant}. It is a typical that risk scorecards are always used in credit acceptance process to optimize and control the risk. Various forms of behavioral scorecards are also used for management of repeat business and also for PD models in Basel RWA (Risk Weighted Assets) calculation~\cite{baselbis}. It is a kind of phenomenon that a list of about 10 account or client characteristics can predict their future behavior, their style of payments and their delinquency.

One can say the trivial fact scorecards are useful and methodology is well known, but on the other hand still credit scoring can be developed and new techniques should be tested. The main problem today is that there is not defined the general testing idea of new methods and techniques, there is no proving method of their correctness. Many good articles are prepared based on one particular case study, on one example of real data coming from one or a few banks~\cite{transitionmatrix},~\cite{huang} and~\cite{sghex}. From a theoretical point of view, even there are presented good results and very correct arguments to suggest choosing one method than another, it is the prove only on that particular data is indicated the difference, but nobody can prove it for other data, nobody can guarantee the correctness for all cases.

There are also other important reasons why real banking data are not available globally and cannot be used by everyone analysts, like legal constraints or too fresh new products with too short data history. These two factors suggest finding a quite another approach for predictive modeling testing in banking usage.

It is a very good idea to start developing two parallel ways: real data and random--simulated data approaches. The second one even cannot replace real data it can be very useful to understand in the better way relations among various factors in data, to imagine a complexity of the process and can be a trial to create more general class of semi-real data.

Let be considered some advantages of randomly generated data:
\begin{itemize}
\item[1.] Today many analysts try to understand and to analyze the last crisis~\cite{stablesupsegment}, among other things they develop methods of indicating risk stable in the time sub-portfolios. Topic is not easy and cannot be solved by typical predictive models based on target variable like in the case of default risk. The notion of stability cannot be defined for every particular account or client, one cannot say that account is stable, only the set of accounts can be tested, so that technique should be developed by quite different method than typical predictive modeling with target variable. It can be formulated by a simple conclusion: the more accounts the more robust stability testing. In the random data generator can be tested various scenario to see and to better understand the problem.
\item[2.] Scoring Challenges or Scoring Olympic Games. From time to time there are organized by different environments contests to find good modelers or to test new techniques. Sometimes data are taken form too real case. Too real means, that some real processes are not predictable, because they are influenced by many immeasurable factors. Even if scoring models are used in practice also in these cases it is not a good idea to use that data for contest. The best solution and the best fairly is to use random data generator process directly predictable.
\item[3.] Reject inference area~\cite{huang}. Still that topic needs development. Random data can be generated also for, in the reality, rejected cases for testing, so it can be used for better estimation of risk on blank areas and better experience.
\item[4.] Today there are two or more techniques of scorecard building~\cite{sasbook}. It needs to make some comparisons, to make some analysis to define recommendations: where and what conditions suggest to use one than another method. The same case can be applied for different variable selection methods.
\item[5.] Product profitability, bad debts and cut-offs. On random data all mentioned notions can be tested and analyst’s experience can be broadened. 
\item[6.] Random data can also be very important factor in the topic of data standardization or the idea of auditing. Let imagine that there are prepared all ready run software tools for MIS (Management Information Systems) and KPI (Key Performance Indicators) reporting on the generic data structure firstly uploaded by random data. Then auditing of all another data will be minimized by only the upload data process.
\end{itemize}

Simulation data are used in many areas, for example it is very useful in research of telecommunication network by the system like OPNET~\cite{opnet}. Also there are developed simulated data in the banking area by ~\cite{simbank} and ~\cite{simgame}.

The simplest retail consumer finance portfolio is the fixed installment loan portfolio. Here process can be simplified by the following assumptions:

\begin{itemize}
\item for all accounts one due date in the middle of the month is defined (every 15th),
\item every client has only one credit,
\item client can pay whole one installment, a few installments or pay nothing, two events only: payment or missing payment,
\item there are measured delinquency on state: end of month by indicated the number of due installments,
\item all customer and account properties are randomly generated by defined proper random distributions,
\item if the number of due installments attain 7 (180 past due days) the process is stopped and account is marked by bad account status, next collection steps are omitted,
\item if number of paid installments attains the number of all installments then the process is stopped and account is marked by closed account status,
\item payments or missing payments are determined by three factors: score calculated on account characteristics, migration matrix and adjustment of that matrix by one cycle time macroeconomic variable,
\item score is calculated for every due installments group separately. In more general case there can be defined different score for every status: due installments 0, 1, ..., and 6. 
\end{itemize}

It is a good circumstance to emphasize that risk management today has very good tools for risk control, even if the crisis has come and was not predicted in the correct way, it could be indicated very quickly. It seems that the best of risk control tools is the migration matrix reporting.

The goal of that paper can be also formulated in the following way: to create random data with the condition to obtain the same results like observed in the reality by typical reporting like migration matrix, flow-rates or roll-rates and vintage or default rates.

\section{Detailed description of data generator \label{detailed}}
\zero
\par

\subsection{The main options}
\par

All data are generated from starting date $T_s$ to ending $T_e$. 


The migration matrix $M_{ij}$ (transition matrix) is defined as a percent of transition after one month from due installments $i$ to due installments $j$.

There is one macro-economic variable dependent only on a time by the formula: $E(m)$, where $m$ is a number of month from $T_s$. It should satisfy the simple condition: $0.01<E(m)<0.9$, because it is used as an adjustment of migration matrix, so it influences on the risk; in some months produces slightly greater one and in some months lower.

\subsection{Production dataset}
\par

The first dataset contains all applications with all available customer characteristics and credit properties.

Customer characteristics (application data):
\begin{itemize}
\item Birthday –- $T_{Birth}$ -- with the distribution $D_{Birth}$ 
\item Income -- $x^a_{Income}$ –- $D_{ Income }$ 
\item Spending -– $x^a_{Spending}$ -- $D_{ Spending }$ 
\item Four nominal characteristics -- $x^a_{Nom_1}, ..., x^a_{Nom_4}$ –- $D_{Nom_1}, D_{Nom_2}, ..., D_{Nom_4} $, in practice they can represent variables like: job category, marital status, home status, education level, or others. 
\item Four interval characteristics –- $x^a_{Int_1}, ..., x^a_{Int_4}$ -- $D_{Int_1}, D_{Int_2}, ..., D_{Int_4} $, represent variables like: job seniority, personal account seniority, number of households, housing spending or others.
\end{itemize}

Credit properties (loan data):
\begin{itemize}
\item Installment amount -- $x^l_{Inst}$ –- with the distribution $D_{ Inst }$ 
\item Number of installments -- $x^l_{N_{inst}}$ -– $D_{ N_{inst} }$ 
\item Loan amount -- $x^l_{Amount}= x^l_{Inst} \cdot x^l_{N_{inst}}$
\item Date of application (year, month) -- $T_{app}$
\item Id of application

\end{itemize}

The number of rows per month is generated based on the distribution $D_{Applications}$.

\subsection{Transaction dataset}
\par

Every row contains the following information (transaction data):
\begin{itemize}
\item Id of application
\item Date of application (year, month) -- $T_{app}$
\item Current month -- $T_{cur}$
\item Number of due installments (number of missing payments) -- $x^t_{n_{due}}$
\item Number of paid installments -- $x^t_{n_{paid}}$

\item Status -- $x^t_{status}$ -– Active (A) –- is still not paid, Closed (C) – is paid, or Bad (B) -- when $x^t_{n_{due}}=7$ 
\item Pay days -- $x^t_{days}$ -- number of days from the interval $[-15,15]$ before or after due date in a current month when payment was done, if there is missing payment, then pay days are also missing.
\end{itemize}

\subsection{Inserting the Production dataset into the Transaction dataset \label{allocation}}
\par
Every month of the Production dataset updates the Transaction dataset with the following formulas:
$$
T_{cur}=T_{app}, \hskip0.5cm
x^t_{n_{due}}=0, \hskip0.5cm
x^t_{n_{paid}}=0, \hskip0.5cm
x^t_{status}=A, \hskip0.5cm
x^t_{days}=0.
$$

It is the process of inserting starting points of new accounts.

\subsection{Analytical Base Table -- ABT dataset}
\par
History of payments for every account is dependent on behavioral data, on behavior of previous payments. It is, of course, the assumption of that data generator.

There are many ideas of behavioral characteristics creation. There are presented the simple methods to consider the last available states and to indicate their evaluations in the time. All data are prepared in ABT datasets, the notion Analytical Base Table is used by SAS Credit Scoring Solution~\cite{saswww}.

Let set current date $T_{cur}$ as a fixed value. Actual states are calculated for that date by the formulas (actual data):
\beqn
x^{act}_{days} =& x^t_{days}+15,  \nonumber \\
x^{act}_{n_{paid}} =& x^t_{n_{paid}},  \nonumber \\
x^{act}_{n_{due}} =& x^t_{n_{due}} , \nonumber \\
x^{act}_{utl} =&  x^t_{n_{paid}} / x^l_{N_{inst}}, \nonumber \\
x^{act}_{dueutl} =&  x^t_{n_{due}} / x^l_{N_{inst}}, \nonumber \\
x^{act}_{age} =& years(T_{Birth},T_{cur}), \nonumber \\
x^{act}_{capacity}  =& ( x^l_{Inst}+ x^a_{Spending}) / x^a_{Income}, \nonumber \\
x^{act}_{dueinc} =&( x^t_{n_{due}} \cdot x^l_{Inst}) / x^a_{Income} , \nonumber \\
x^{act}_{loaninc} =&  x^l_{Amount} / x^a_{Income}, \nonumber \\
x^{act}_{seniority} =&  T_{cur}-T_{app}+1, \nonumber
\eeqn

where $years()$ calculates the difference between two dates in years.

Let consider two time series of pay days and due installments for the last 11 months from fixed current date by the formulas:

\beqn
x^{act}_{days} (m) =& x^{act}_{days}(T_{cur} - m),  \nonumber \\
x^{act}_{n_{due}} (m) =& x^{act}_{n_{due}} (T_{cur} - m) , \nonumber
\eeqn

where $m=0,1, ...,11$.  

The characteristics indicated the evaluation in the time can be calculated by the formulas:

If every elements of time series for the last $t$-months are available then (behavioral data):
\beqn
x^{beh}_{days} (t) =& (\sum_{m=0}^{t-1} x^{act}_{days}(m))/ t,  \nonumber \\
x^{beh}_{n_{due}} (t) =& (\sum_{m=0}^{t-1} x^{act}_{n_{due}}(m)) / t , \nonumber
\eeqn
where $t=3, 6, 9, 12$.

If not all elements of time series are available then (missing imputation formulas):
\beqn
x^{beh}_{days} (t) =& 15,  \nonumber \\
x^{beh}_{n_{due}} (t) =& 2. \label{missinginput}
\eeqn
In other words behavioral variables represent average states for last 3, 6, 9 or 12 months. Without any problem user can add many other variables by replacing average statistic by another like MAX, MIN or other.

\subsection{Migration matrix adjustment}
\par 
Macro-economic variable $E(m)$ influenses on the migration matrix by the formula:
\begin{displaymath}
M^{adj}_{ij} = \left\{ \begin{array}{ll}
M_{ij}(1-E(m)) & \textrm{for} \hskip0.5cm
 j \le i, \\
M_{ij} & \textrm{for}\hskip0.5cm
 j > i+1,\\
M_{ij}+\sum_{k=0}^{i} E(m)M_{ik} & \textrm{for}\hskip0.5cm
 j=i+1.
\end{array} \right.
\end{displaymath}

\subsection{Iteration step}
\par 
That step is running to generate next month of transactions, from $T_{cur}$ to $T_{cur}+1$. In every month some accounts are new, then the Transaction dataset is only updated by the ideas described in the subsection \ref{allocation}. 
Some accounts change the status by the formula:

\begin{displaymath}
x^t_{status}= \left\{ \begin{array}{ll}
C & \textrm{when} \hskip0.5cm
 x^{act}_{n_{paid}} = x^l_{N_{inst}}, \\
B & \textrm{when}\hskip0.5cm
 x^{act}_{n_{due}} = 7,
\end{array} \right.
\end{displaymath}
and these accounts are not continued in next months.

For other active accounts in the next month there are generated events: payment or missing payment. It is based on two scorings:

\beqn
Score_{Main}=&\sum_{\alpha} \beta^a_{\alpha} x^a_{\alpha}
+\sum_{\gamma} \beta^l_{\gamma } x^l_{\gamma }
+\sum_{\delta} \beta^{act}_{\delta } x^{act}_{\delta } \nonumber \\
+&\sum_{\eta} \sum_t \beta^{beh}_{\eta } (t) x^{beh}_{\eta } (t)
+ \beta_r \varepsilon
+ \beta_0, \label{scoremain} \\
Score_{Cycle}=&\sum_{\alpha} \phi^a_{\alpha} x^a_{\alpha}
+\sum_{\gamma} \phi^l_{\gamma } x^l_{\gamma }
+\sum_{\delta} \phi^{act}_{\delta } x^{act}_{\delta } \nonumber \\
+&\sum_{\eta} \sum_t \phi^{beh}_{\eta } (t) x^{beh}_{\eta } (t)
+\phi_r \epsilon
+ \phi_0 \label{scorecycle},
\eeqn
where
$t=3, 6, 9, 12$, \hskip0.3cm
$\alpha=Income, Spending, Nom_1,...,Nom_4, Int_1,..., Int_4$, \hskip0.3cm
$\gamma=Inst, N_{Inst}, Amount$, 
\hskip0.3cm
$\eta=days,n_{due}$, \hskip0.3cm
$\delta=days, n_{paid}, n_{due}, utl, dueutl,$ $age, capacity, dueinc, loaninc, seniority$, \hskip0.3cm
$\varepsilon$ and $\epsilon$ are taken from the standardized normal distribution $N$.

Let consider the following migration matrix:
\begin{displaymath}
M^{act}_{ij}= \left\{ \begin{array}{ll}
M^{adj}_{ij} & \textrm{when} \hskip0.5cm Score_{Cycle}\le \textrm{Cutoff}
, \\
M_{ij} & \textrm{when} \hskip0.5cm Score_{Cycle}> \textrm{Cutoff},

\end{array} \right.
\end{displaymath}
where $\textrm{Cutoff}$ is another parameter like all $\beta$s and $\phi$s.

For fixed $T_{cur}$ and fixed $x^{act}_{n_{due}}=i$ all active accounts can be segmented by $Score_{Main}$ to satisfy the same proportions like appropriate elements of migration matrix $M^{act}_{ij}$:
the first group $g=0$ by the highest scores has share equaled to $M^{act}_{i0}$, the second $g=1$ has share $M^{act}_{i1}$, ..., and the last group $g=7$ share -- $M^{act}_{i7}$.

For particular account assigned to the group $g$ payment is done in month $T_{cur}+1$ when $g\le i$, in other case payment is missing.

For missing payment Transaction dataset is updated by the following information:
$$
x^t_{n_{paid}}=x^{act}_{n_{paid}},
$$
$$
x^t_{n_{due}}=g,
$$
$$
x^t_{days}=\textrm{Missing}.
$$

For payment by formulas:
$$
x^t_{n_{paid}}=\min(x^{act}_{n_{paid}}+ x^{act}_{n_{due}}-g+1, x^l_{N_{inst}}),
$$
$$
x^t_{n_{due}}=g,
$$
and
$x^t_{days}$ are generated from the distribution $D_{days}$.

Described steps are repeated for all months between $T_s$ and $T_e$.

\subsection{Default definition}
\par 

The Default is a typical credit scoring and Basel II notion. Every account from the observation point $T_{cur}$ is tested during the outcome period equals 3, 6, 9 and 12 months. During that time there is analyzed maximal number of due installments, exactly: 
$$
\textrm{MAX}= \textrm{MAX}_{m=0}^{t-1} (x^{act}_{n_{due}} (T_{cur}+m)),
$$ where $t=3, 6, 9, 12$. Dependently on value $\textrm{MAX}$ are defined three values of default statuses $\textrm{Default}_t$:

{\bf Good:} When $\textrm{MAX} \le 1$ or during the outcome period was $x^t_{status}=C$.

{\bf Bad:} When $\textrm{MAX}>3$ or during the outcome period $x^t_{status}=B$. In the case $t=3$ when $\textrm{MAX}>2$.

{\bf Indeterminate:} for other cases.

Existing of Indeterminate status can be questionable. In some analysis only two statuses are preferable, for example in Basel II. It is also a good topic for father research which can be solved due to data generator described in this paper.

\subsection{Portfolio segmentation and risk measures}
\par

Typically credit scoring is used for the control of the following sub--portfolios or processes:

{\bf Acceptance process -- APP portfolio:} It is the set of all starting points of credits, where it is decided which one are accepted or rejected. Acceptance sub--portfolio is defined as the set of rows of Transaction dataset with the condition: $T_{cur}=T_{app}$. Every account belongs to that set only ones.

{\bf Cross--up sell process -- BEH portfolio:} It is the set of all accounts with the longer history than 2 months and in the good condition (without delinquency). Cross--up sell or Behavioral sub--portfolio is defined as the set of rows of Transaction dataset with the condition: $x^{act}_{seniority}>2$ and $x^{act}_{n_{due}}=0$. Every account can belongs to that set many times.

{\bf Collection process -- COL portfolio:} It is the set of all accounts with the delinquency, but at the beginning of the collection process. Collection sub--portfolio is defined as the set of rows of Transaction dataset with the condition: $x^{act}_{n_{due}}=1$. Every account can belongs to that set many times.

For every mentioned sub--portfolio one can calculates and tests risk measures called bad rates defined as the share of {\bf Bad} statuses for every observation points and outcome periods.

Definitions of mentioned sub--portfolios in the reality can be more complex, here are suggested the simplest versions for father analysis of cases studies presented in the section \ref{twocases}.

\section{General theory}
\zero
\par

\subsection{The main assumption and definition}
\par

{\bf Definition.} The layout 
$$
(T_s, T_e, M_{ij}, E(m), 
\beta^a_{\alpha},
\beta^l_{\gamma }, 
\beta^{act}_{\delta }, 
\beta^{beh}_{\eta } (t), 
\beta_r, 
\beta_0,
$$
$$
\phi^a_{\alpha},
\phi^l_{\gamma }, 
\phi^{act}_{\delta }, 
\phi^{beh}_{\eta } (t), 
\phi_r,
\phi_0,
\varepsilon,
\epsilon,
D_{Birth}, D_\alpha, D_\gamma, D_{Applications}, D_{days},
\textrm{Cutoff})
$$ with the all rules and symbols, relations and processes described in the section \ref{detailed} is called {\bf The Retail Consumer Finance Data Generator in the case of fixed installment loans} with the nick name {\bf RCFDG}.

{\bf Theorem -- assumption.} Every consumer finance portfolio with the fixed installment loans can be estimated by the {\bf RCFDG}.

The proof of that theorem can be always done in the correct way due to parts:
$\beta_r \varepsilon$ and $\phi_r \epsilon$ in the formulas \ref{scoremain} and \ref{scorecycle}. From the empirical point of view credit scoring is always used in portfolio control, so mentioned theorem is correct, but problem is with the goodness of fit. Up to now theory is too early to define a good measures of fit, however it is a proper starting point in the next development of the general theory of consumer finance portfolios.

The similar ideas and researches are presented in~\cite{transitionmatrix}.

\subsection{Open questions}
\par 

The next steps probably would be concentrated on:
\begin{itemize}
\item Finding the correct goodness of fit statistics measuring the distance between the real consumer finance portfolio and {\bf RCFDG}. Also it should be tested the property of that statistics.
\item Analyzing the additional constraints to satisfy for example properties like: the predictive power, measured for example by Gini~\cite{wp14}, of characteristic 
$x^{beh}_{days}(3)$ on $\textrm{Default}_6$ should be equaled to $40\%$.
\item Creating more general case with all collection processes, more than one credit per customer, more than one macro-economic factors and other detailed issues.
\item Analyzing of various existing real consumer finance portfolios and finding the set of parameters describing each of them. Then there can be developed the theory of principal component analysis (PCA) of all consumer finance portfolios in the particular country or in the world.
\item Defining the generalization of the notion of consumer finance portfolio contains almost all properties of real portfolios. 
\item Using that generalized notion in researches on the development of scoring methods to use that notion as a general idea of method proving. For example the theorem: {\it Scoring models build on $\textrm{Default}_3$ and on $\textrm{Default}_{12}$ produce the same results} could be solved by the additional condition: betas for $t=3$ and for $t=12$ should be similar. It is very probable that many future researches will discover many properties and relations among betas, coefficients of the migration matrix and their consequences.
\end{itemize}

\section{Two case studies \label{twocases}}
\zero
\par

\subsection{Common parameters}
\par

All random numbers are based on two typical random generators: 
uniform $U$ and standardized normal $N$ distributions, in details: the distribution $U$ returns a number from the interval $(0,1)$ with the equal probability.

All common coefficients are the following:
$T_s=$ 1970.01 (January 1970), $T_e=$ 1976.12 (December 1976), 

\begin{displaymath}
M_{ij}= \left\lbrack \begin{array}{ccccccccc}
    &   j=0&  j=1&  j=2&  j=3&  j=4&  j=5&  j=6&  j=7 \\
i=0 & 0.850&0.150&0.000&0.000&0.000&0.000&0.000&0.000 \\
i=1 & 0.250&0.450&0.300&0.000&0.000&0.000&0.000&0.000 \\
i=2 & 0.040&0.240&0.190&0.530&0.000&0.000&0.000&0.000 \\
i=3 & 0.005&0.025&0.080&0.100&0.790&0.000&0.000&0.000 \\
i=4 & 0.000&0.000&0.010&0.080&0.090&0.820&0.000&0.000 \\
i=5 & 0.000&0.000&0.000&0.000&0.020&0.030&0.950&0.000 \\
i=6 & 0.000&0.000&0.000&0.000&0.000&0.010&0.010&0.980 \\
\end{array} \right\rbrack,
\end{displaymath}
$E(m)= 0.01+(1.5+\sin((5 \cdot \pi \cdot m)/(T_e-T_s))+N/5)/8$,
$D_{Applications}=300 \cdot 30 \cdot (1+N/20)$,
if $T_{app}$ is December then $D_{Applications}=D_{Applications} \cdot 1.2$. To define $D_{Birth}$ first is defined distribution of age:
$D_{Age}=((75-18)\cdot (N+4)/7 + 10 + 20\cdot U)$ if $Age>75$ then $Age=75$, if $Age<18$ then $Age=18$. $D_{Birth}=T_{app}-D_{Age}\cdot 365.5$,
$D_{Income}=int((10000-500)/40\cdot 10 \cdot abs(N)+500)$,
$D_{Inst}=int(Income\cdot abs(N)/4)$,
$D_{Spending}=int(Income\cdot abs(N)/4)$,
$D_{N_{Inst}}=int(30\cdot abs(N)/4+6)$ if $N_{Inst}<6$ then $N_{Inst}=6$, 
$D_{Nom_i}=int(5\cdot abs(N))$ and $D_{Int_i}=10\cdot U$, for $i=1, 2, 3, 4$,
if $x^{act}_{n_{due}}<2$ then $D_{days}=-int(15\cdot (abs(N)/4))$
else $D_{days}=int(15\cdot (N/4))$,
where $int()$ and $abs()$ are integer value and absolute value suitable. 

To avoid scale or unit problem for every individual variable it is suggested to make a simple standardization step for ABT table for every $T_{cur}$ before score calculation. That idea is quite realistic, because even some customers are good payers in the crisis time they can also have more problems, so general condition of the current month can influence on all customers. On the other hand to present interesting two cases is decided to standardize variables by the global parameters.

Scoring formula for $Score_{Main}$ is calculated based on the table \ref{scoring}, namely:
$$
Score_{Main} = \sum_{index=1}^{28} \beta (x-\mu) / \sigma .
$$

All beta coefficients could be recalculated without standardization step, but in that case it would be more difficult to interpret them. By a simple study of the table \ref{scoring} it can be indicated that the most significant variables have absolute value equals 6.

\begin{table}
\begin{center}
\caption{Scoring formula for $Score_{Main}$.}
\label{scoring}
\centerline{\scriptsize
\begin{tabular}{|l|l|l|l|l|}
 \hline
  & & & & \\
  Index & $x$ -- variable & $\mu$ & $\sigma$ & $\beta$ \\
  & & & & \\
 \hline
1& $x^a_{Nom_1}     $ & 3.5 & 3 & 1 \\
2& $x^a_{Nom_2}     $ & 3.5 & 3 & 2 \\
3& $x^a_{Nom_3}     $ & 3.5 & 3 & 1 \\
4& $x^a_{Nom_4}     $ & 3.5 & 3 & 3 \\
5& $x^a_{Int_1}     $ & 5 & 2.89 & 1 \\
6& $x^a_{Int_2}     $ & 5 & 2.89 & -4 \\
7& $x^a_{Int_3}     $ & 5 & 2.89 & 1 \\
8& $x^a_{Int_4}     $ & 5 & 2.89 & -2 \\
9& $x^{act}_{days}  $ & 13 & 2.42 & -5 \\
10& $x^{act}_{utl}   $ & 0.36 & 0.28 & -4 \\
11& $x^{act}_{dueutl} $ & 0.12 & 0.2 & -6 \\
12& $x^{act}_{n_due}  $ & 1.3 & 2 & -2 \\
13& $x^{act}_{age}     $ & 53 & 9.9 & 4 \\
14& $x^{act}_{capacity}$ & 0.4 & 0.21 & -2 \\
15& $x^{act}_{dueinc}  $ & 0.3 & 0.6 & -1 \\
16& $x^{act}_{loaninc} $ & 2.4 & 2.1 & -2 \\
17& $x^a_{Income}      $ & 2395 & 1431 & 2 \\
18& $x^l_{Amount}      $ & 5741 & 6804 & -1 \\
19& $x^l_{N_{inst}}$ & 12.3 & 4.63 & -4 \\
20& $x^{beh}_{n_{due}}(3)$ & 1.4 & 1.6 & -4 \\
21& $x^{beh}_{days}   (3)$ & 14.15 & 1.4 & -6 \\
22& $x^{beh}_{n_{due}}(6)$ & 1.6 & 1.13 & -5 \\
23& $x^{beh}_{days}   (6)$ & 14.57 & 1.02 & -6  \\
24& $x^{beh}_{n_{due}}(9)$ & 1.78 & 0.75 & -5 \\
25& $x^{beh}_{days}   (9)$ & 14.78 & 0.72 & -6 \\
26& $x^{beh}_{n_{due}}(12)$ & 1.89 & 0.48 & -5 \\
27& $x^{beh}_{days}   (12)$ & 14.91 & 0.49 & -6  \\
28& $\varepsilon$ & 0 & 0.02916 & 1 \\
\hline
\end{tabular}
}
\end{center}
\end{table}

\subsection{The first case study -- {\bf unstable application characteristic} -- APP}
\par 

In that case it is assumed that only customers with low income can be influenced by a crisis. Application characteristic income in that data generator is a stable variable during the time, and the migration matrix is adjusted by the macro--economic $E(m)$ only for cases:
$$
x^a_{Income}<1800.
$$

Presented relation without any problem can be transformed into the general form 
\ref{scorecycle}. 

\subsection{The second case study -- {\bf unstable behavioral characteristic} -- BEH}
\par 
Here the condition for migration matrix adjustment is the following:
$$
x^{beh}_{n_{due}}(6)>0 \hskip0.5cm \textrm{and} \hskip0.5cm x^{act}_{seniority}>6,
$$
the rule for the seniority variable is added to not adjust accounts with missing imputation based on \ref{missinginput}. That case presents situation when crisis has an impact on customers who had some delinquency during their last 6 months. 

\subsection{Stability problem}

\par 

Let be considered the typical scoring models building process, for example on behavioral sub--portfolio. Because two cases are based on two variables one application and one behavioral let be considered only the set of these two variables.
To indicate strong instability models they are analyzed with the target variable $\textrm{Default}_9$.

Every variable is segmented or binned for a few attributes described in the tables \ref{binningAPP} and \ref{binningBEH}.

\begin{figure}
\caption{Risk measures on $\textrm{Default}_9$ on attribute 2 of variable $x^a_{Income}$ for two cases APP and BEH.}
\label{income}
\begin{center}
\includegraphics[angle=-90, width=0.8\textwidth]{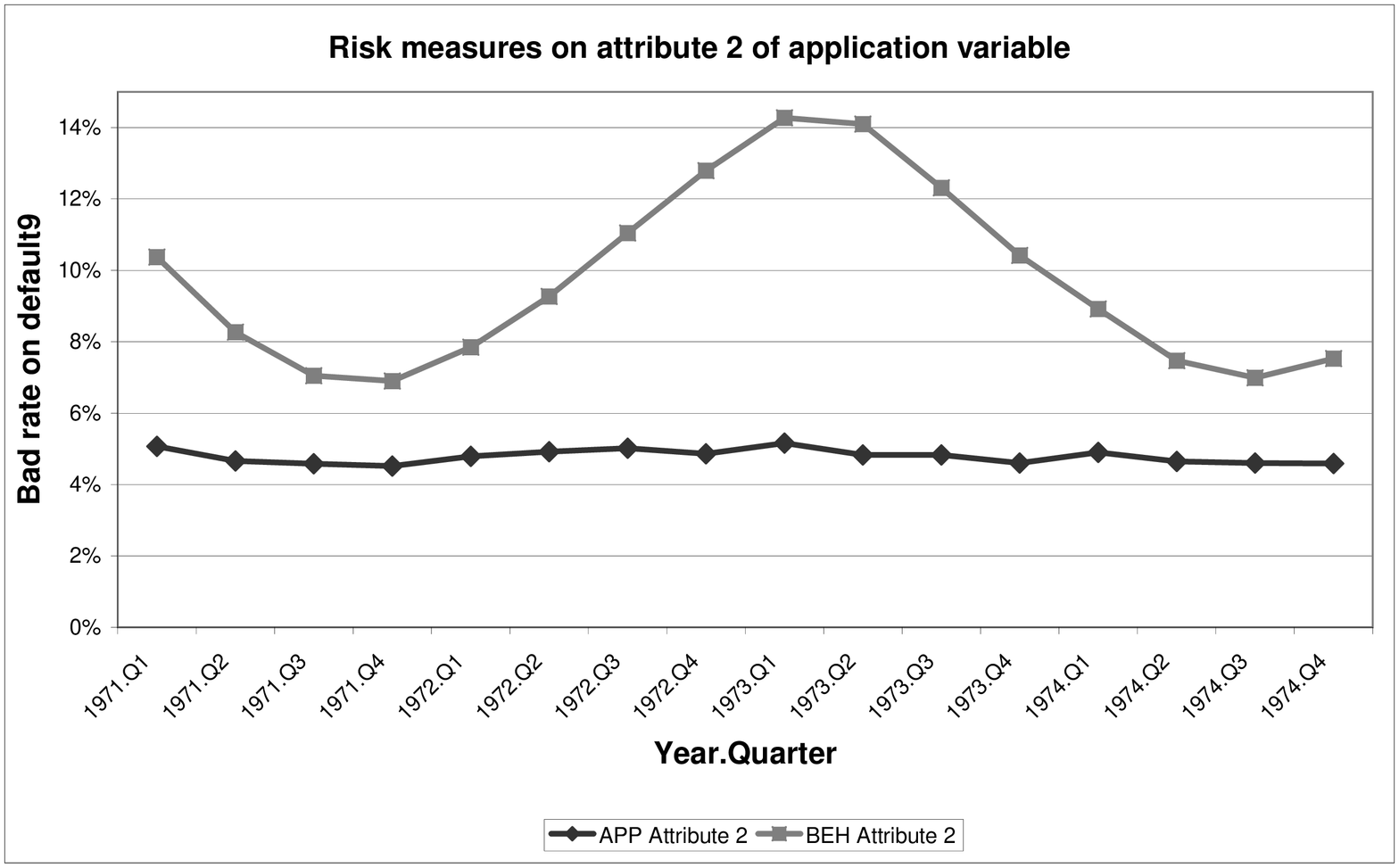}
\end{center}
\end{figure}

\begin{figure}
\caption{Risk measures on $\textrm{Default}_9$ on attributes of variable $x^a_{Income}$ for the case APP.}
\label{income_app}
\begin{center}
\includegraphics[angle=-90, width=0.8\textwidth]{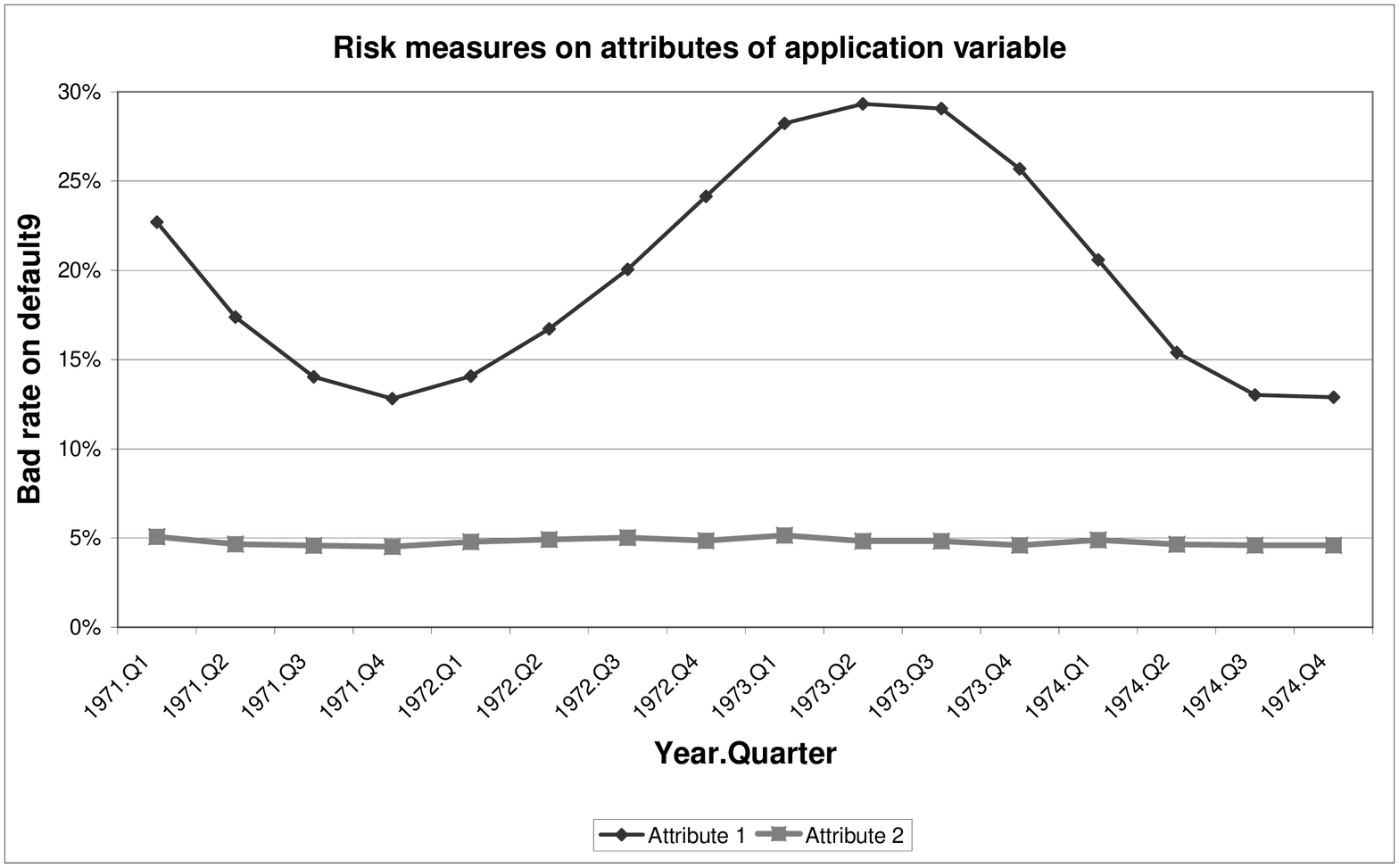}
\end{center}
\end{figure}

\begin{figure}
\caption{Risk measures on $\textrm{Default}_9$ on attributes of variable $x^a_{Income}$ for the case BEH.}
\label{income_beh}

\begin{center}
\includegraphics[angle=-90, width=0.8\textwidth]{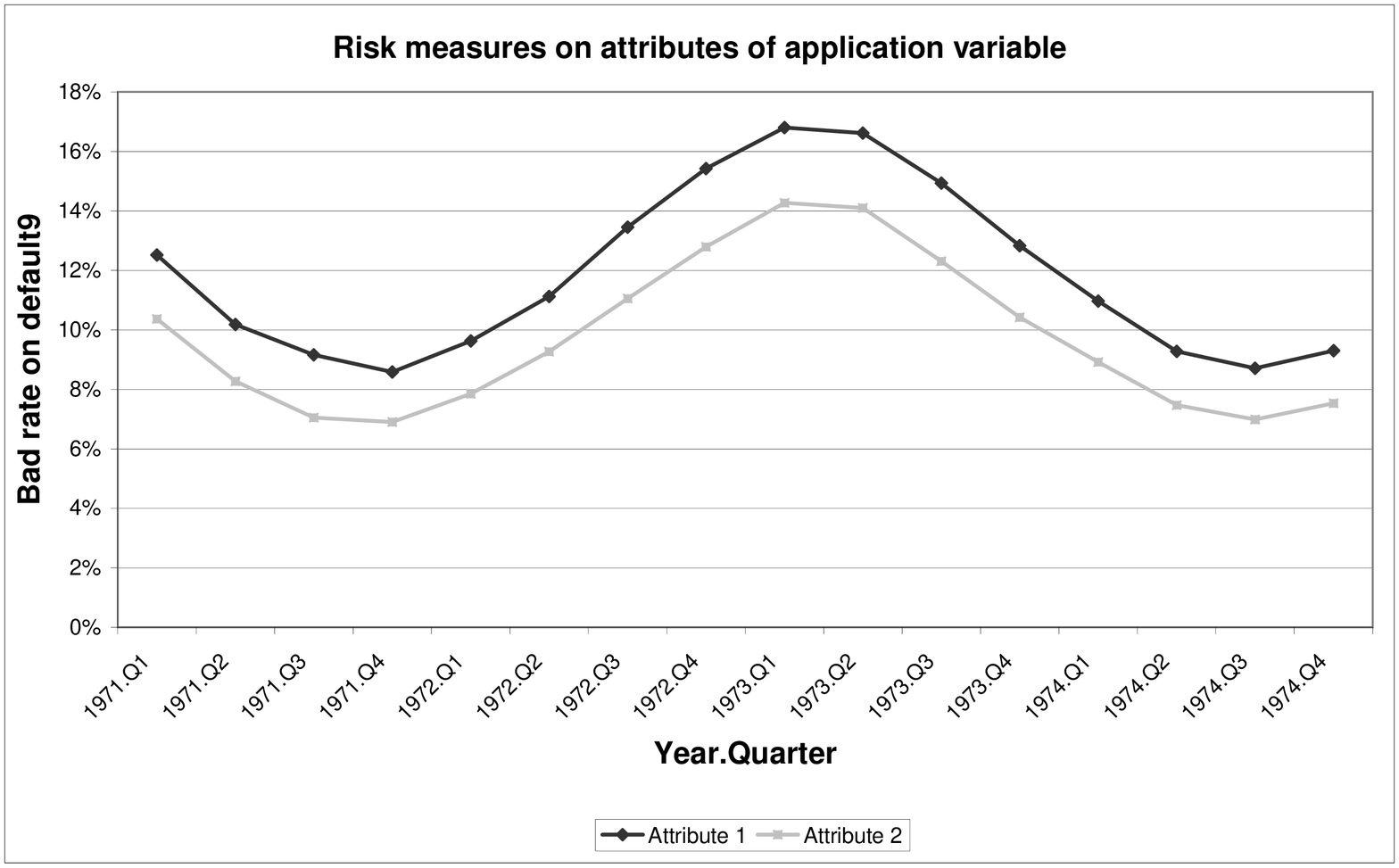}
\end{center}
\end{figure}

In the case of unstable application variable (APP) by studying the figure \ref{income_app} can be confirmed, what is expected, that attribute 2 is very stable during the time and accounts from that group are not quite sensible for crisis changes. In opposite attribute 1 is very unstable. The same groups in the case of unstable behavioral variable (BEH) are both unstable, see the figure \ref{income_beh}. The same group, accounts from attribute 2, are presented on figure \ref{income} for both cases to indicate in a better scale that APP case can really choose accounts not sensitive on the crisis. Even data generator is simplicity of the real data, that conclusion is very useful. Some application data can be profitable in risk management to indicate sub--segments with stable risk in the time.

\begin{figure}
\caption{Risk measures on $\textrm{Default}_9$ on attribute 3 of variable $x^{beh}_{n_{due}}(6)$ for two cases APP and BEH.}
\label{agr6_mean_due}
\begin{center}
\includegraphics[angle=-90, width=0.8\textwidth]{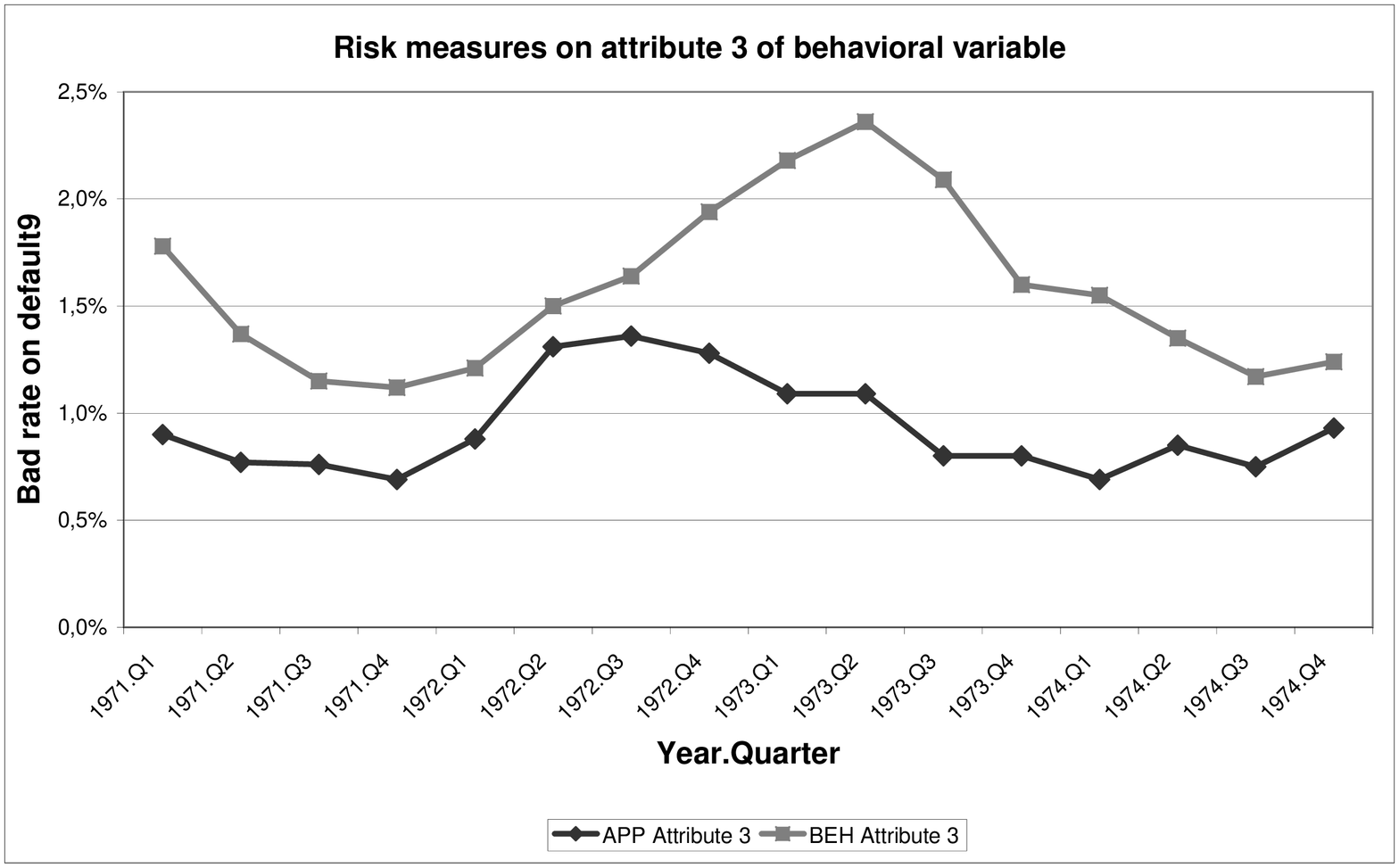}
\end{center}
\end{figure}

\begin{figure}
\caption{Risk measures on $\textrm{Default}_9$ on attributes of variable $x^{beh}_{n_{due}}(6)$ for the case APP.}
\label{agr6_mean_due_app}
\begin{center}
\includegraphics[angle=-90, width=0.8\textwidth]{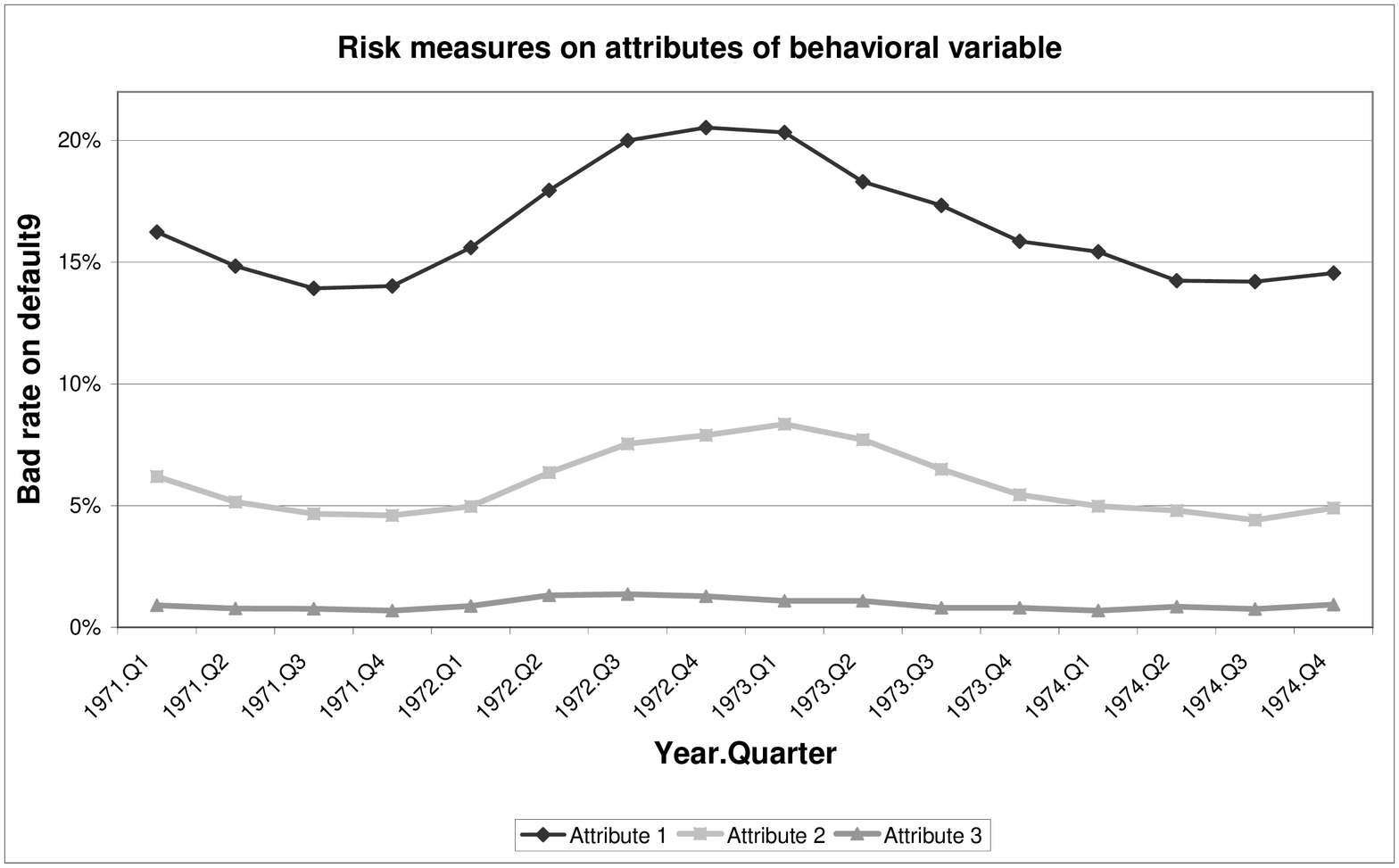}
\end{center}
\end{figure}

\begin{figure}
\caption{Risk measures on $\textrm{Default}_9$ on attributes of variable $x^{beh}_{n_{due}}(6)$ for the case BEH.}
\label{agr6_mean_due_beh}
\begin{center}
\includegraphics[angle=-90, width=0.8\textwidth]{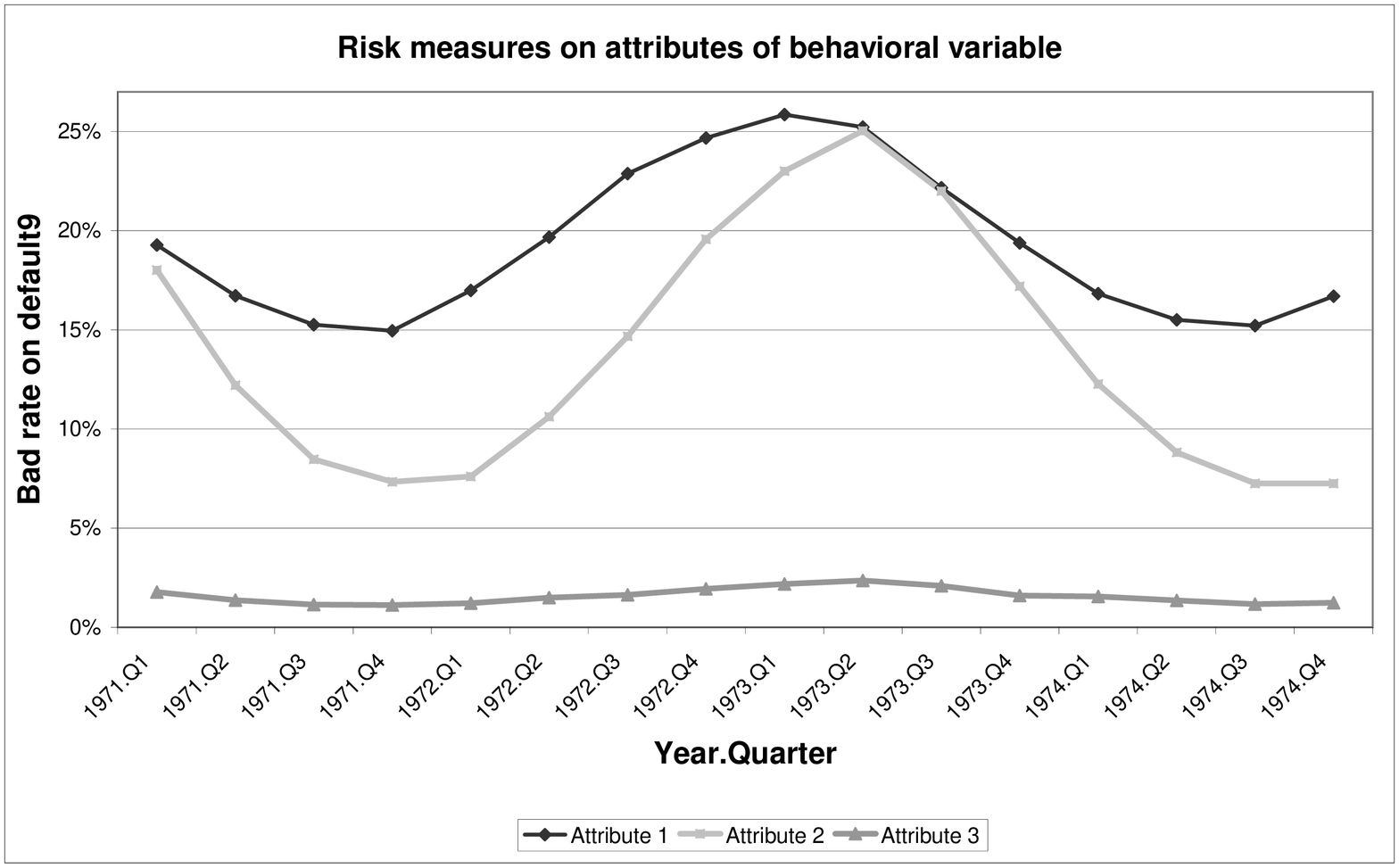}
\end{center}
\end{figure}

Not the same conclusions can be formulated for behavioral variable $x^{beh}_{n_{due}}(6)$. On the figure \ref{agr6_mean_due_app} there are presented risk evolutions for three attributes of that variable. All of them are not stable. The most stable attribute is with the number 3. Also for the case BEH that attribute is not stable, see the figure \ref{agr6_mean_due_beh}. To be sure of that there are also presented on the figure \ref{agr6_mean_due} only attributes 3 for both cases. Every reader can say that both cases have unstable risk. Even in the case BEH the attribute 3 is expected to have a stable risk, due to the rule for migration matrix adjustment, expectation has failed. The reason comes from the correct understanding of the process. Typical scoring approach is based on the principal idea that historical information up to the observation point is able to predict behavior during the outcome period. Up to the observation point account did not have any delinquency so the variable $x^{beh}_{n_{due}}(6)=0$. After that point in the next months account can have due installments. It can be adjusted by the macro--economic variable and on the end that group can become unstable.

The mentioned idea is very important for father research of the crisis. It should be emphasized that typical scoring methods used on three types of sub--portfolios: APP, BEH and COL cannot discover in the correct way the rule of crisis adjustment and cannot indicate some sub--segments stable in the time. Of course scoring can be also used just like in that paper for prediction of migration states; to be very clear, not for default statuses prediction but for transition prediction. The best method is probably the survival analysis~\cite{survival} or~\cite{dynamicsurvival} with time covariates (time dependent variables), where in natural way there is indicated the factor of being better or worse payer in the correct time, namely in the typical scoring model the factor is considered but only up to the observation point. In the survival model however it can be also taken into the account after that observation point, so in the more realistic way.

There are made many other cases of data generators with more complex rule for $Score_{Cycle}$. If there are taken together both types of variables: application and behavioral the case is too complicated and unstable property exists everywhere. In that case is not possible to find stable factor. That conclusion is also very important for crisis analysis, because it describes the nature of crisis: if it is a strong event and it has an impact on both types of characteristics behavioral and application -- it is and risk management can try to find some sub--segments only more stable then others or with maximal risk not exceeded the expected boundary. 

\begin{table}
\begin{center}
\caption{Simple binning for two variables in the case APP.}
\label{binningAPP}

{\scriptsize

\begin{tabular}{ c c c c c c }
Characteristic & Attribute & Condition & Bad rate &	Population & Gini \\

& number & & on $\textrm{Default}_9$ & percent & on $\textrm{Default}_9$\\

\hline
$x^{beh}_{n_{due}}(6)$ & 

\begin{tabular}{c}
1 \\ 2 \\ 3
\end{tabular} &

\begin{tabular}{c}
$x^{act}_{seniority}<6$  \\ $x^{beh}_{n_{due}}(6)>0$ 
and $x^{act}_{seniority}\ge 6$  \\ otherwise
\end{tabular} &

\begin{tabular}{ r @{.} l  }
16&77\% \\
6&48\% \\
1&07\% \\
\end{tabular} &
 
\begin{tabular}{ r @{.} l  }
37&09\% \\
22&49\% \\
40&42\% \\
\end{tabular} &

 51.34\%
\\
\hline


$x^a_{Income}$ & 

\begin{tabular}{c}
1 \\ 2 
\end{tabular} &

\begin{tabular}{c}
$x^a_{Income}<1800$ \\ $x^a_{Income}\ge 1800$ \end{tabular} &

\begin{tabular}{ r @{.} l  }
20&11\% \\
4&72\% \\
\end{tabular} &

\begin{tabular}{ r @{.} l  }
18&32\% \\
81&68\% \\
\end{tabular} &

 36.29\%

\end{tabular}
}
\end{center}
\end{table}

\begin{table}
\begin{center}
\caption{Simple binning for two variables in the case BEH.}
\label{binningBEH}

{\scriptsize

\begin{tabular}{ c c c c c c }
Characteristic & Attribute & Condition & Bad rate &	Population & Gini \\

& number & & on $\textrm{Default}_9$ & percent & on $\textrm{Default}_9$\\

\hline
$x^{beh}_{n_{due}}(6)$ & 

\begin{tabular}{c}
1 \\ 2 \\ 3
\end{tabular} &

\begin{tabular}{c}
$x^{act}_{seniority}<6$  \\ $x^{beh}_{n_{due}}(6)>0$ 
and $x^{act}_{seniority}\ge 6$  \\ otherwise
\end{tabular} &

\begin{tabular}{ r @{.} l  }
19&49\% \\
14&04\% \\
1&74\% \\
\end{tabular} &
 
\begin{tabular}{ r @{.} l  }
40&05\% \\
16&52\% \\
43&43\% \\
\end{tabular} &

 46.54\%
\\
\hline


$x^a_{Income}$ & 

\begin{tabular}{c}
1 \\ 2 
\end{tabular} &

\begin{tabular}{c}
$x^a_{Income}<1800$ \\ $x^a_{Income}\ge 1800$ \end{tabular} &

\begin{tabular}{ r @{.} l  }
12&09\% \\
10&09\% \\
\end{tabular} &
 
\begin{tabular}{ r @{.} l  }
39&49\% \\
60&51\% \\
\end{tabular} &

 5.04\%

\end{tabular}
}
\end{center}
\end{table}

\subsection{Various types of risk measures}

\par 

\begin{figure}
\caption{Risk measures on $\textrm{Default}_9$ comparison on sub--portfolios: APP, BEH and COL and also with one flow rate $M_{23}$.}
\label{risk_beh}
\begin{center}
\includegraphics[angle=-90, width=0.8\textwidth]{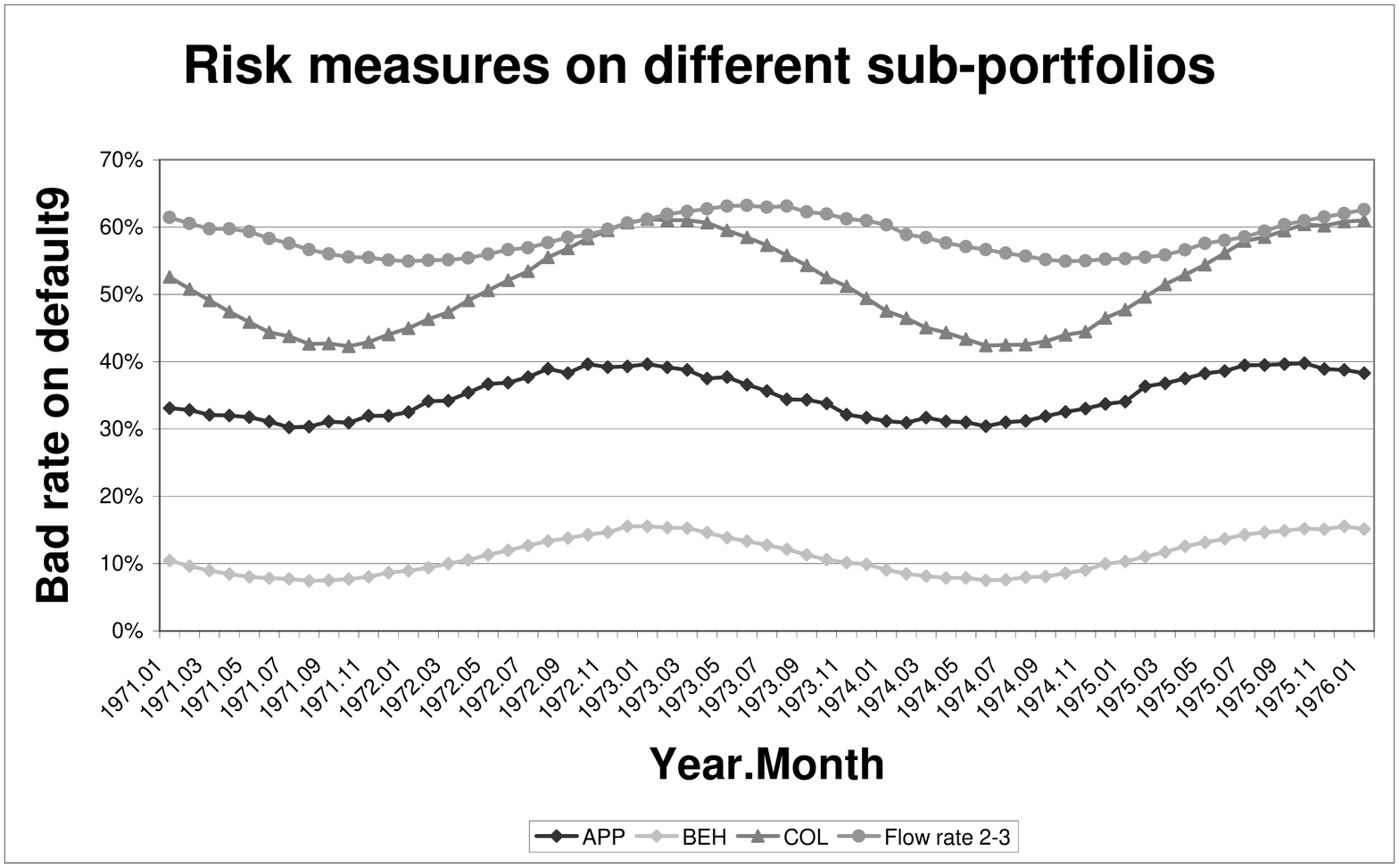}
\end{center}
\end{figure}

Let be defined that crisis is a time where risk is the highest. The most popular reporting for risk management is based on bad rates, vintage and flow rates. The figure \ref{risk_beh} presents bad rates for three different sub--portfolios application, behavioral and collection. There is presented also one flow rate. There is a simple conclusion that crisis does not occur in the same time. Some curves indicate local maximum of risk earlier than others. The difference in the time is significant and can be almost 6 months, so it is very important to remember what kind of reports can indicate a crisis as quickly as possible. It should be emphasized that bad rates reports present, by the standard way, the evaluation of risk by observation points and a crisis time can occur between observation point and the end of outcome period. It seems that flow rates reports precise the crisis time in better way.


\subsection{Implementation}
\par

All data were prepared by the SAS System~\cite{saswww} by manual codes written in SAS 4GL used units: Base SAS and SAS/STAT.
For the case of unstable behavioral variable -- BEH: Production dataset has 779~993 rows (about 90MB) and Transaction dataset -- 8~969~413 rows (about 400MB). Total time of calculation per one case takes about 4 hours.

\section{Conclusions}
\zero
\par

Even if data are generated by random--simulated process, which is not realistic, the conclusions give the possibility to better understand the nature of the crisis.

The banking data generator is a new hope for researching to find the proving method of comparisons of various credit scoring techniques. It is probable that in the future many random generated data will become the new repository for testing and comparisons.

In the first case -- unstable application variable like income is possible to split portfolio for two parts: stable and unstable during the time. For the second case unstable -- behavioral characteristic the task is more complicated and it is not possible to split in the same way. Some sub--segments can have better stability but always they fluctuate. Moreover if a crisis is impacted by many factors both from application form customer characteristics and from a customer behavioral together it is very difficult to indicate these factors and the crisis in reports is everywhere.

Generated data are very useful for various analysis and researches. There are many rows, many bad default statuses, so analyst can make many good exercises to improve his experience.


\newcommand{\byauthors}[1]{#1 }
\newcommand{\journal}[1]{ #1 }
\newcommand{\reftitle}[1]{{\it #1} }
\newcommand{\volumin}[1]{{\bf #1} }
\newcommand{\eref}{.}
\newcommand{\refyear}[1]{(#1), }


\bibliographystyle{plunsrt}        

\bibliography{bib}

\end{document}